\documentclass[submission,copyright,creativecommons]{eptcs}
\providecommand{\event}{Formal Methods for Autonomous Systems} % Name of the event you are submitting to

\usepackage{iftex}

\usepackage[T1]{fontenc}
\usepackage{graphicx}
\usepackage{listings}
\usepackage{subcaption}
\usepackage{rotating}
\usepackage{amsmath}
\usepackage{centernot}
\usepackage{hyperref}
\usepackage{fretish}
\usepackage{pgfgantt}
\usepackage{array}
\usepackage{enumitem}
\usepackage{pdflscape}

\usepackage{comment}

\usepackage[linesnumbered, ruled]{algorithm2e}

\usepackage{listings}
\usepackage{xcolor}

\definecolor{codegreen}{rgb}{0,0.6,0}
\definecolor{codegray}{rgb}{0.5,0.5,0.5}
\definecolor{codepurple}{rgb}{0.58,0,0.82}
\definecolor{backcolour}{rgb}{0.95,0.95,0.92}

\lstdefinestyle{mystyle}{
    backgroundcolor=\color{backcolour},   
    commentstyle=\color{codegreen},
    keywordstyle=\color{magenta},
    numberstyle=\tiny\color{codegray},
    stringstyle=\color{codepurple},
    basicstyle=\ttfamily\footnotesize,
    breakatwhitespace=false,         
    breaklines=true,                 
    captionpos=b,                    
    keepspaces=true,                 
    numbers=left,                    
    numbersep=5pt,                  
    showspaces=false,                
    showstringspaces=false,
    showtabs=false,                  
    tabsize=2
}

\newcommand{\gwendolen}{{\sc Gwendolen}}
\newcommand{\mcapl}{{\sc MCAPL}}
\newcommand{\ajpf}{{\sc AJPF}}
\newcommand{\java}{{\sc Java}}
\newcommand{\jpf}{{\sc JPF}}

\usepackage{todonotes}

\title{Safe-ROS: An Architecture for Autonomous Robots in Safety-Critical Domains}

\def\manchester{
  \institute{Department of Computer Science\\
             The University of Manchester\\
             Manchester, UK}
}

\author{Diana C. Benjumea \qquad
Marie Farrell \qquad
Louise A. Dennis \qquad
\institute{University of Manchester\\
Manchester, UK}
\email{
  \{diana.benjumeahernandez,  marie.farrell, louise.dennis\}@manchester.ac.uk
}}

\def\titlerunning{Safe-ROS: An Architecture for Autonomous Robots}
\def\authorrunning{D.C. Benjumea, M. Farrell \& L.A. Dennis }

\begin{document}

\maketitle

\begin{abstract}

Deploying autonomous robots in safety-critical domains requires architectures that ensure operational effectiveness and safety compliance. In this paper, we contribute the Safe-ROS architecture for developing reliable and verifiable autonomous robots in such domains. It features two distinct subsystems: (1) an intelligent control system that is responsible for normal/routine operations, and (2) a Safety System consisting of Safety Instrumented Functions (SIFs) that provide formally verifiable independent oversight. We demonstrate Safe-ROS on an AgileX Scout Mini robot performing autonomous inspection in a nuclear environment. One safety requirement is selected and instantiated as a SIF. To support verification, we implement the SIF as a cognitive agent, programmed to stop the robot whenever it detects that it is too close to an obstacle. We verify that the agent meets the safety requirement and integrate it into the autonomous inspection. This integration is also verified, and the full deployment is validated in a Gazebo simulation, and lab testing. We evaluate this architecture in the context of the UK nuclear sector, where safety and regulation are crucial aspects of deployment. Success criteria include the development of a formal property from the safety requirement, implementation, and verification of the SIF, and the integration of the SIF into the operational robotic autonomous system. Our results demonstrate that the  Safe-ROS architecture can provide safety verifiable oversight while deploying autonomous robots in safety-critical domains, offering a robust framework that can be extended to additional requirements and various applications.


\end{abstract}

\section{Introduction}
\label{Sec:Introduction}

When deploying autonomous systems (i.e., that make decisions without human intervention) in \textit{safety-critical} domains, such as nuclear~\cite{LuckcuckWorkshopReport,abeywickrama2025autonomy} or aerospace~\cite{moy2013testing}, various standards and regulators dictate that we must provide strong guarantees that the system is doing what it is expected to do in terms of safety requirements~\cite{IEC61508,BS61513,IEC61513:2011,farrell2021evolution}. Autonomous systems offer significant benefits in these domains, including removing humans from harm in nuclear environments \cite{nagatani2013emergency}. But since critical decisions would now be made by software rather than human operators, ensuring correctness and reliability through robust verification is necessary to help build trust in the autonomous decision-making components that drive these systems.

The Robot Operating System (ROS)~\cite{quigley2009ros} is a commonly used middleware for developing robotic systems. It is used in various industries, offering benefits such as flexibility, modularity, and integration of diverse packages often developed in different programming languages. ROS enables communication between software components, facilitating complex robot behaviours through modular design. However, when applying ROS in safety-critical domains, significant challenges arise, particularly regarding the verification of safety properties and system trustworthi{\-}ness \cite{luckcuck2019formal}. Key questions include: 
\begin{itemize}[noitemsep,nosep]
    \item[\textbf{Q1.}] What framework can be applied, at the ROS application layer, to prevent the occurrence of unsafe outcomes when using probabilistic models?
    \item[\textbf{Q2.}] How can we provide formal guarantees that the overall system meets its safety requirements, behaves as expected, and is safe to use?
    \item[\textbf{Q3.}] And finally, how can we trace the entire process, from eliciting requirements, through analysis and formalization, to code integration, verification, and beyond—across the system life cycle?
\end{itemize}

In this paper, we contribute the Safe-ROS architecture. The idea of Safe-ROS comes from mitigating the lack of verifiable safety decision-making in ROS applications. Traditional C++ and Python packages distributed by the ROS community are widely used across industries, but they often lack formal verification and may not be transparent to the system programmer. For this work, the packages responsible for core middleware functionality (e.g., communication between nodes and messages flow) remain unverifiable, Safe-ROS focuses specifically on the application layer, where unverified probabilistic methods directly influence decision-making. Our contribution involves an architecture that integrates an agent that can be verified against safety requirements and that detects and responds to hazardous situations, ultimately bringing the robot to a safe state. In our architecture, this agent acts as the Safety Instrumented Function (SIF). A SIF is a vital component of safety systems in various industries, including nuclear~\cite{NS-TAST-GD-094,IEC61508}, chemical processing~\cite{IEC61511,HSESIS,BS6739}, and oil and gas~\cite{IEC61511}. It is designed to prevent or mitigate hazardous events by taking specific actions when certain conditions are met. 

Various approaches have been developed to address safety in robotics, such as emergency stop mechanisms, safety barrier functions ~\cite{ISO10218-1-2025,ames2016control}, and similar safeguards. More proactive control strategies have been introduced~\cite{comi2019modelling}, but these often lack formal guarantees, which limits their suitability for safety-critical applications. This leaves an important gap in ensuring reliable system performance, particularly under unexpected conditions. In this work, we are instantiating a system architecture that enhances ROS-based autonomous navigation by integrating a Belief, Desire, and Intentions (BDI) agent. This approach has been previously explored in~\cite{dal2022developing}, and we additionally provide a verification process that allows us to produce strong guarantees that the system is free from unexpected and dangerous behaviours. This presents a valuable contribution to deploying autonomous systems safely in critical domains.

The remainder of this paper is structured as follows: \textbf{Section}~\ref{Sec:PrototypeInstantiationOfTheArchitecture} presents the core technical work, focusing on implementation,  verification and validation of a nuclear inspection use case. We describe the implementation of a motion controller using ROS for autonomous inspection with the AgileX Scout Mini Robot (\textbf{Subsection}~\ref{Sec:SafetyRelatedAutonomousSystem}), and the Safety Instrumented Function (SIF), a BDI agent that is programmed to maintain a safe distance from obstacles (\textbf{Subsection}~\ref{Sec:SafetyInstrumentedFunction}). We then explain the integration of these subsystems (\textbf{Subsection}~\ref{Sec:SRASandSSIntegration}). \textbf{Subsection}~\ref{sec:verification} presents the formal requirements elicitation process and the integration of verification and validation techniques, including model checking in MCAPL, deductive verification with Dafny, and testing. \textbf{Section}~\ref{Sec:Evaluation} evaluates how our approach provides safety guarantees, enables reliable ROS package integration, and supports system traceability. Finally, \textbf{Section}~\ref{Sec:RelatedWork} presents related work, and \textbf{Section}~\ref{Sec:DiscussionAndFurtherWork} concludes the paper with discussion and directions for future research. Next, we provide some background prerequisites for our work and discuss related approaches.

\section{Background}
\begin{figure} [t]
    \centerline{\includegraphics[scale=0.21]{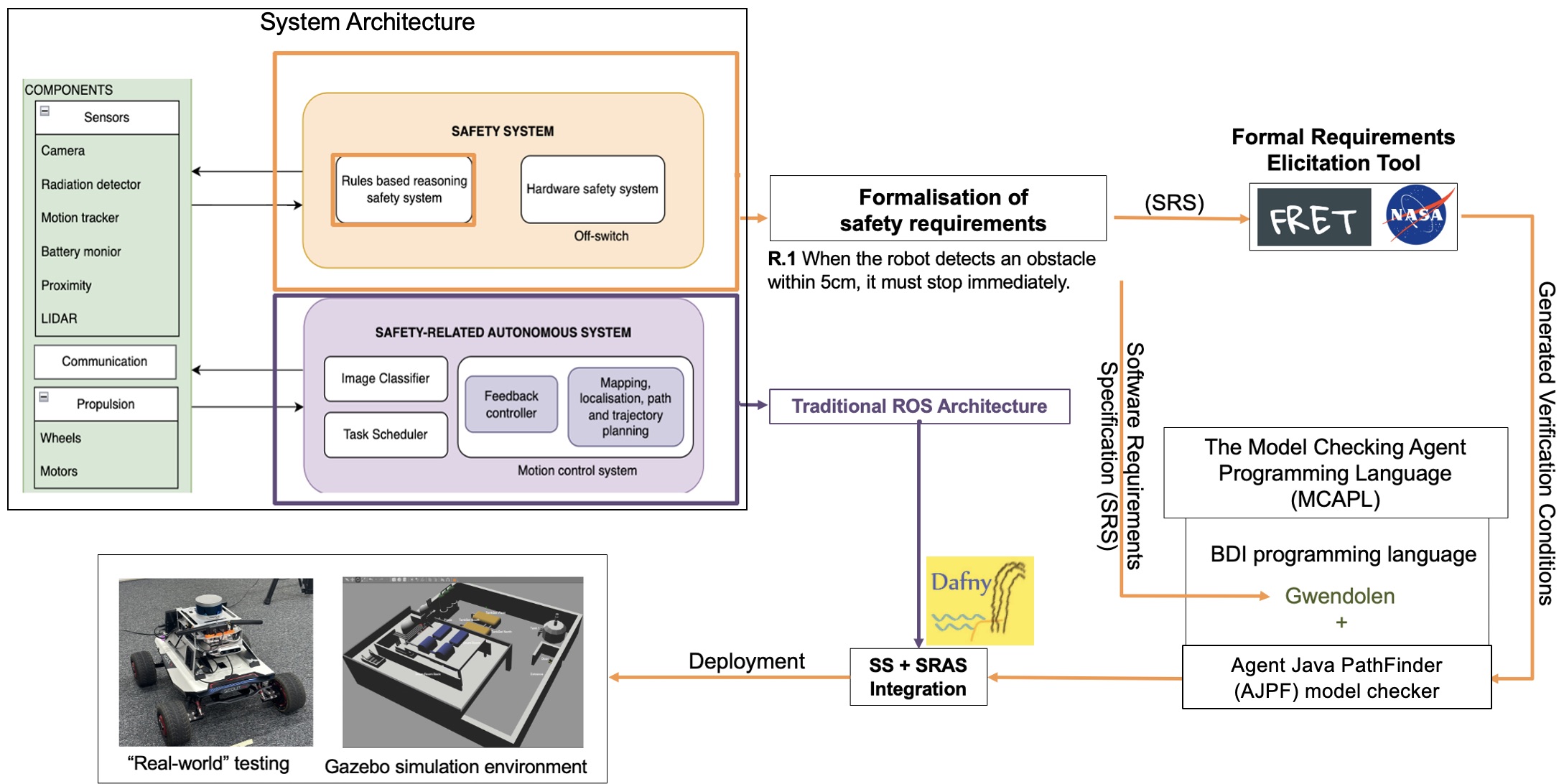}}
    \caption{Research methodology: Our approach to incorporating formal methods throughout our safety architecture.}
    \label{Fig:methodology}
\end{figure}

\textbf{Approach Overview:} \textbf{Figure}~\ref{Fig:methodology} shows our approach, which builds on prior work~\cite{anderson2023autonomous,benjumea2024formalising}. The system architecture in the top-left comprises various physical components commonly found in autonomous robotic systems. It features two key subsystems. The \emph{Safety Related Autonomous System (SRAS)} uses autonomous control technologies such as neural networks and feedback controllers for routine operation and is overseen by a \emph{Safety System} implementing Safety Instrumented Functions (SIFs), which provide reversionary control and ensure safety-critical requirements while retaining some intelligence. This approach traces requirements from elicitation and hazard analysis through design to implementation and verifies that the SIFs enforce them. In this work, the SRAS is implemented as a ROS-based motion controller, and we refer to the architectural artefact as Safe-ROS. 

Our implementation of Safe-ROS follows a workflow (shown in \textbf{Figure}~\ref{Fig:methodology}) that connects safety requirements engineering with formal verification and system integration. Safety requirements are first elicited in structured natural language using the Formal Requirements Elicitation Tool (FRET)~\cite{pressburger2022fret}, which supports the structured definition of verifiable safety goals. These requirements are automatically  translated into temporal logic specifications using FRET and verified using model checking and theorem proving techniques. Specifically, the behaviour of the SIF is modelled as a BDI agent~\cite{rao1999foundations} and verified using (\ajpf)~\cite{dennis2018mcapl}, ensuring logical consistency and correctness of its decision-making process. Additional deductive verification in Dafny~\cite{Rustan2010Dafny} provides assurance of software correctness for the integration of the SIF and the SRAS.

\textit{Scope of applicability.} Safe-ROS is a general architectural pattern for developing safety-wrapped autonomous systems in domains where external safety supervision is feasible and formally verifiable. It is best suited for cyber-physical robots operating in bounded, structured environments, such as ground robots performing inspection in nuclear settings with well-defined hazards and operational constraints. The framework focuses on enforcing safety rules at the application level (e.g., maintaining safe distances, speed limits, or initiating emergency stops) rather than verifying low-level sensor processing, ROS middleware communication, or overall system safety.

\textbf{Motivation for formal verification.} Autonomous robotic systems often rely on probabilistic and/or unverified components, such as perception, path planning, or learning-based controllers, whose behaviour cannot be guaranteed under all conditions. In Safe-ROS, we target formal verification specifically at the Safety Instrumented Function (SIF), the component responsible for monitoring and intervening on the Safety-Related Autonomous System (SRAS). By formally specifying and model checking the SIF using \ajpf\ and verifying key safety functions in \textit{Dafny}, we ensure that runtime-enforceable safety properties are preserved, even if other SRAS components behave unpredictably.

\textbf{The Robot Operating System} (ROS)~\cite{quigley2009ros} is a widely adopted middleware framework that simplifies the integration and orchestration of robotic software components. Before the development of ROS, programming robots required specialized interfaces and significant engineering effort to enable communication between software modules. ROS introduced a standardized architecture based on nodes, topics, and services, making it significantly easier to develop and scale complex robotic systems~\cite{ROS-originStory}. Today, ROS is a fundamental platform for academic and industrial robotics projects, supported by a large and active community. According to the 2024 ROS Metrics Report~\cite{ROS-metricReport}, the ecosystem comprises over 3,000 packages, is used by more than 1,250 companies, and sees millions of downloads each month.

\textbf{The Model Checking Agent Programming Language} (\mcapl) framework contains the tools and interpreters required to build rational agent programming languages and also integrates the Agent Java PathFinder (\ajpf) model checker~\cite{dennis2018mcapl}.
\mcapl\ formally verifies BDI-based agent programs by exhaustively exploring all possible behaviours to ensure correctness. Its combination of execution and verification tools makes it well-suited for developing reliable, safety-critical autonomous systems. 

\textbf{Gwendolen} is a BDI (Belief, Desire and Intentions) programming language~\cite{rao1999foundations} designed for creating \textit{verifiable agents}; cognitive agents, to be precise. The concept of a cognitive agent originates from philosophy and cognitive science and is defined as an entity that makes decisions based on clear, explicit reasons and should be able to explain its choices if needed~\cite{dennis2023verifiable}. The motivation for using a BDI programming language to build the SIF comes from the idea of encoding safety requirements as a set of beliefs or rules of the system. This approach implements a rules-based system that provides transparency in the system’s logical decision-making process, making it suitable for formal verification.

When designing an autonomous safety rules-based reasoning system, especially for critical applications like UK nuclear safety, it is crucial to select a language that offers robustness, reliability, expressiveness, and strong verification capabilities. After reviewing the literature on agent programming languages~\cite{dennis2017gwendolen,bordini2007programming,busetta1999jack,dastani20082apl,hindriks2009programming}, \gwendolen\ was chosen primarily due to its strong support for formal verification. One key advantage is that model checking tools are embedded within the framework, simplifying the implementation process. Although its user community is relatively small, our collaboration with the \gwendolen\ research group has provided valuable support and expertise during development.

\textbf{Formal Verification} is the process of assessing whether a specification, expressed in logic, is satisfied by a given system description. This can be done through deductive verification, where a system description $\psi_S$ is shown to logically imply a property $\varphi$ (i.e., $\vdash \psi_S \Rightarrow \varphi$), or via algorithmic approaches such as model checking, where a model $M$ is automatically checked to satisfy $\varphi$ (i.e., $M \models \varphi$)~\cite{fetzer1988program,de1979social,boyer1983proof,clarke1997model}. Traditional model checking verifies properties over an abstract system model. In contrast, \emph{model checking of programs} directly verifies properties over the program's actual execution paths~\cite{visser2003model}. This is possible in languages like \java, where tools such as Java Pathfinder (\jpf) use a modified virtual machine to explore all possible program behaviours, enabling verification without an abstracted model.

\textbf{Agent Java Pathfinder} (\ajpf)~\cite{dennis2018mcapl} extends \jpf\ to verify agent-based systems, particularly those using Belief-Desire-Intention (BDI) architectures. It checks properties across all possible executions, considering control flow and the agent's reasoning. \ajpf\ has been shown as a practical platform for automated verification of multi-agent programs~\cite{bordini2008automated}. Here, we use \ajpf\ to model check a BDI agent acting as the SIF in our system, ensuring behaviours meet formal safety requirements.

\textbf{Dafny}~\cite{Rustan2010Dafny} is a language-based formal verification system used in research~\cite{farrell2021using, farrell2022formal} and industry~\cite{chakarov2022better}. Most programming languages were not designed with verification in mind, but in order to facilitate writing proofs through an automatic process, specialized verification-aware languages have been developed. Examples include WhyML~\cite{filliatre2013why3}, F*~\cite{rastogi2023proof}, and SPARK~\cite{carre1990spark}. For this work, we are using Dafny. The Dafny verifier has the benefit of running continuously in supported Integrated Development Environments (IDEs). Dafny programs are translated into the Boogie intermediate verification language and verified using Z3. Whenever it cannot automatically verify a proof obligation, it flags it as an error, similar to how a word processor highlights spelling mistakes. Dafny integrates formal verification into a programming environment, providing formal guarantees that a program satisfies its specifications under all possible executions~\cite{Rustan2010Dafny}. A wide range of resources support learning Dafny, including the official manual~\cite{leino2021dafny}, interactive tutorial~\cite{nipkow2012getting}, Reynolds’s online tutorial collection~\cite{conor_dafny}, and textbook~\cite{leino2023program}.

\textbf{Nuclear Inspection} is central to operating, maintaining, and decommissioning facilities. In the UK, the Nuclear Decommissioning Authority’s (NDA) Business Plan 2025--2028~\cite{nda2025draft} targets decommissioning 17 legacy sites, including Sellafield, which accounts for approximately 85\% of the UK’s nuclear waste. A major challenge is conducting inspections safely in environments that are hazardous to humans due to radiation, contamination, and/or structural instability. Recent efforts ~\cite{saveliev2024multi,mitchell2023lessons} highlight the growing relevance of robotic deployment in nuclear environments and recognise that this domain demands much higher levels of safety and reliability to prevent harm to humans, assets, and critical infrastructure. To support the safe design and validation of such systems, realistic simulation environments have been developed~\cite{watson2014remote}, and robotic platforms have been reinforced to enable autonomous monitoring and characterisation of indoor nuclear facilities~\cite{nouri2023carma}. However, their deployment remains limited, and from the software perspective, there is a need for verifiable, independent, and diverse Safety Systems~\cite{NS-TAST-GD-046}.

Our case study focuses on routine inspections at a UK nuclear site. The hazard analysis and requirements elicitation process for an autonomous robot carrying out this operation was previously presented in~\cite{benjumea2024formalising}. A ground-based autonomous robot was selected as the optimal solution, balancing mission effectiveness and operational safety. We use the AgileX Scout Mini (shown in \textbf{Figure}~\ref{Fig:methodology}), a four-wheel-drive commercial robot with a ROS development kit, industrial control, LiDAR, and multiple sensors. Its capabilities support motion control, communication, navigation, and map building~\cite{generationrobots2021scout}.

\section{A Prototype Instantiation of the Architecture}
\label{Sec:PrototypeInstantiationOfTheArchitecture}

We have implemented the proposed architecture (illustrated in \textbf{Figure}~\ref{Fig:methodology}) on an AgileX Scout Mini robot, performing an inspection task. In this section, we discuss the implementation of the Safety-related Autonomous System and the Safety System (which implements the Safety Instrumented Function). We also describe the integration of these two subsystems, along with the validation and verification process.

\textbf{Code Availability: }This work provides an open-source artifact -  \href{https://github.com/dianabenjumea/Safe-ROS}{GitHub project}\footnote{GitHub project: https://github.com/dianabenjumea/Safe-ROS}. Containing the full implementation of the Safe-ROS architecture, including the SRAS controller, Safety System, verification tools, and simulation environment. This supports reproducibility and facilitates further research.

\subsection{Safety-Related Autonomous System (SRAS)} 
\label{Sec:SafetyRelatedAutonomousSystem}

\begin{algorithm}[t]
\caption{SRAS Workflow for Autonomous Navigation in ROS}
\label{alg:sras-workflow}
{\scriptsize
\KwData{
    \begin{tabular}{@{}ll@{}}
\underline{Topics} & \underline{Nodes} \\
\texttt{/velodyne\_points}: LiDAR & \texttt{pointcloud\_to\_laserscan}: Converts 3D LiDAR to 2D scans \\
    \texttt{/scan}: 2D scan & \texttt{rf2o\_laser\_odometry}: Estimates planar motion from scans \\
    \texttt{/odom}: odometry & \texttt{amcl}: Monte Carlo Localization on the map \\
    \texttt{/map}: map & \texttt{map\_server}: Loads and serves the static map \\
    \texttt{/goal}: goal & \texttt{simple\_navigation\_goals}: Publishes navigation goals \\
    \texttt{/cmd\_vel}: velocity commands & \texttt{move\_base}: Plans and executes navigation paths \\
     & \texttt{robot\_state\_publisher}, \texttt{joint\_state\_publisher}: TFs and joint states \\
    \end{tabular}
}
\vspace{0.5em}
\KwResult{Autonomous navigation using LiDAR and predefined goals}

\SetKwBlock{Begin}{begin}{end}
\Begin{
  $\texttt{/scan} \gets \texttt{pointcloud\_to\_laserscan}(\texttt{/velodyne\_points})$\;
  $\texttt{/odom} \gets \texttt{rf2o\_laser\_odometry}(\texttt{/scan})$\;
  $\texttt{/map} \gets \texttt{map\_server}()$\;
  $pose \gets \texttt{amcl}(\texttt{/scan}, \texttt{/map}, \texttt{/initialpose})$\;
  $\texttt{/goal} \gets \texttt{simple\_navigation\_goals}()$\;
  $\texttt{robot\_state\_publisher}()$, $\texttt{joint\_state\_publisher}()$\;
  $\texttt{/cmd\_vel} \gets \texttt{move\_base}(\texttt{/odom}, \texttt{/scan}, \texttt{/map}, \texttt{/goal})$\;
  \Return \texttt{/cmd\_vel}
}
}
\end{algorithm}

\begin{figure}[t]
    \centerline{\includegraphics[scale=0.48]{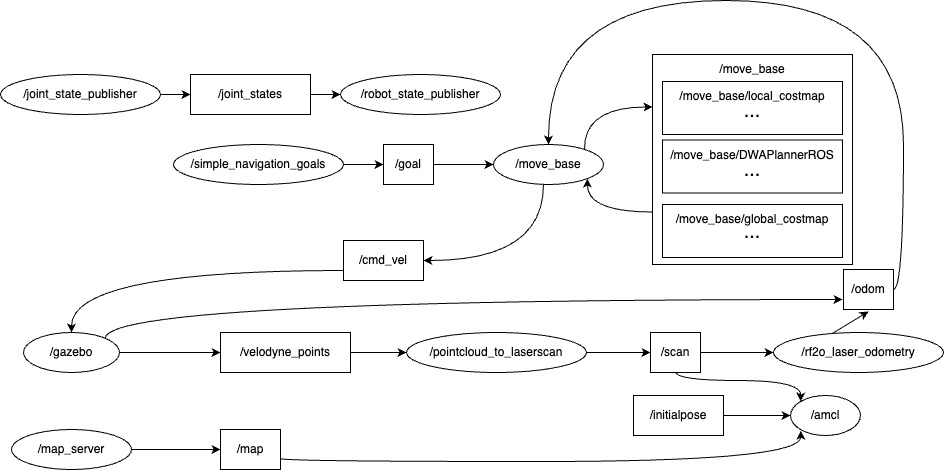}}
    \caption{ROS node graph of the autonomous system. Ovals represent nodes, rectangles denote topics, and arrows indicate message flow. Showing which nodes publish to or subscribe to which topics. 
    }
    \label{Fig:rosgraph}
\end{figure}

For the Safety-Related Autonomous System (SRAS) we implement a motion controller using the ROS Noetic navigation stack, which is in line with previous work~\cite{anderson2023autonomous}. It is designed to operate within a simulated environment to test the autonomous navigation capabilities that are relevant to obstacle avoidance. The simulation environment is created in Gazebo using a nuclear waste storage room model provided by the University of Manchester \cite{manchester_dataset}. For the robotic platform, we adapted the AgileX simulation model from~\cite{agilex_sim}, and incorporated a LiDAR sensor via a dedicated Gazebo plugin.

The overall workflow for autonomous navigation is summarized in \textbf{Algorithm}~\ref{alg:sras-workflow}, and \textbf{Figure}~\ref{Fig:rosgraph} illustrates the interconnection between nodes and topics within the ROS Noetic framework, as deployed in the Gazebo-based simulation environment. At the core of the system is the perception pipeline, which begins with the Gazebo simulator publishing sensor data, including 3D point clouds from a simulated LiDAR via the \texttt{/velodyne\_points} topic. These point clouds are processed by the \texttt{pointcloud\_to\_laserscan} node~\cite{pointcloud_to_laserscan}, which converts them into 2D laser scans published on the \texttt{/scan} topic (line~2 of \textbf{Algorithm}~\ref{alg:sras-workflow}). This 2D scan data serves as a common input for both localization and odometry estimation processes. 

The \texttt{rf2o\_laser\_odometry} node uses laser scan to estimate planar motion~\cite{rf2o_laser_odometry} and publishes odometry data on the \texttt{/odom} topic (line~3). Meanwhile, the static map is loaded via the \texttt{map\_server} node (line~4). Localization is handled by the \texttt{amcl} node~\cite{amcl}, which subscribes to the \texttt{/scan}, \texttt{/initialpose}, and \texttt{/map} topics to estimate the robot’s pose on a static map provided by the \texttt{map\_server} (line~5). This estimated pose aligns the robot’s perceived position with the known environment.

Navigation goals are defined and sent by a custom package, \texttt{simple\_navigation\_goals} (line~6), and the robot’s kinematic state is broadcast by \texttt{robot\_state\_publisher} and \texttt{joint\_state\_publisher} nodes, ensuring accurate relative positioning of all of the sensors and actuators (line~7). 

Navigation is orchestrated by the \texttt{move\_base} node~\cite{move_base}, which plays a central role in path planning and execution. It receives input from the localization and odometry systems via the \texttt{/odom}, \texttt{/scan}, and \texttt{/map} topics, and uses both global and local costmaps to plan collision-free paths. Navigation goals are issued through the \texttt{/goal} topic, and corresponding velocity commands are published on the \texttt{/cmd\_vel} topic to drive the robot toward its target (line~8). Internally, \texttt{move\_base} manages the DWA (Dynamic Window Approach) planner and multiple costmap layers, all shown in the ROS graph (\textbf{Figure}~\ref{Fig:rosgraph}). These probabilistic methods can occasionally produce unsafe motions due to sensor noise, localization errors, or unexpected environmental conditions. Safe-ROS mitigates this via a verified Safety Instrumented Function (SIF) that, in this work, stops the robot when it approaches an obstacle, demonstrating the architecture’s feasibility and potential for more complex safety properties. Next, we describe the implementation of our Safety System to ensure collision avoidance.

\subsection{Safety System (SS)}
\label{Sec:SafetyInstrumentedFunction}

\begin{figure}[t]

\begin{lstlisting} [language=Java, caption=Environment logic (code snippet), label=lst:env, basicstyle=\scriptsize\ttfamily ]
... 
private static final String CONTROL_TOPIC = "/gwendolen_control";
private static final String CONTROL_TYPE = "std_msgs/Bool";
...
bridge.subscribe(SubscriptionRequestMsg.generate("/scan")
    .setType("sensor_msgs/LaserScan"),(data, rep) -> handleLaserScanData(data));
...
private void handleLaserScanData(JsonNode data) {
    double minValue = extractMinRange(data.get("msg").get("ranges"));
    if (minValue < 0.05) {
        addPercept(new Literal("too_close"));
    }
}
public Unifier executeAction(String agName, Action act) {
    if (act.getFunctor().equals("stop_moving")) {
        publishStopSignal();
    }
    return new Unifier();
}
private void publishStopSignal() {
    Publisher control = new Publisher(CONTROL_TOPIC, CONTROL_TYPE, bridge);
    control.publish(new PrimitiveMsg<>(true));
}
...
\end{lstlisting}
\end{figure}

\begin{figure} [t]
\begin{lstlisting} [language=prolog, caption=Safety Instrumented Function, label=lst:sif, basicstyle=\scriptsize\ttfamily]
GWENDOLEN
:name: agilex_agent

:Plans:
+too_close: {True} <-
    stop_moving,
    +stopped;  
\end{lstlisting}
\end{figure}

For the SS we adopted the \gwendolen\ programming language~\cite{dennis2017gwendolen} from the \mcapl\ framework~\cite{dennis2018mcapl,dennis2023verifiable}. \mcapl\ enables us to verify \gwendolen\ programs using the \ajpf\ model checker.  Furthermore, we can integrate a \gwendolen\ program into a ROS based system using the ROS-A framework~\cite{cardoso2020interface}. At the moment, the SS consists of a single Safety Instrumented Function (SIF). We identified one safety requirement from~\cite{benjumea2024formalising} to be instantiated in the SIF.  This requirement is that {\bf the robot shall maintain a safe distance from obstacles}, and was instantiated as:

\begin{enumerate}[nosep]
    \item[\textbf{R1:}]\emph{When the robot detects that an obstacle is within 5cm of it, then it must stop immediately.}
\end{enumerate}

This safety requirement is implemented within a \gwendolen\ program that continuously monitors the robot’s environment using simulated LiDAR data (\texttt{/scan} topic) from Gazebo. 

The environment, summarized in \textbf{Listing}~\ref{lst:env}, is implemented in \java . It establishes a WebSocket connection to ROS, subscribes to the LiDAR topic, and continuously processes incoming messages (lines 5–7). Sensor readings are parsed to identify the minimum range value; if this falls below a threshold, the environment generates a \texttt{too\_close} percept and inserts it into the agent’s belief base. This behavior is handled by the \texttt{handleLaserScanData} method, implemented in lines 8–13. The environment also manages the execution of actions that are issued by the agent, such as publishing to the \texttt{/gwendolen\_control} topic to command the robot to stop, as implemented in the \texttt{executeAction} method (lines 14–19).

This resulted in a small program shown in \textbf{Listing}~\ref{lst:sif}. A \gwendolen\ agent, \texttt{agilex\_agent} (line 2), reasons over environmental percepts from LiDAR data. It has a single plan: if it receives a \texttt{too\_close} percept when an obstacle is detected (line 5), it triggers the \texttt{stop\_moving} action (line 6), which publishes a control signal to stop the robot and adds a percept indicating the robot has stopped (line 7).

\subsection{SRAS \& SS Integration: Using \texttt{java\_rosbridge} and an Orchestrator Node. }
\label{Sec:SRASandSSIntegration}

In a final deployment of the Safe-ROS architecture, the SRAS and SS should be independent, diverse, and segregated. However, given resource availability and aiming to present a proof of concept of this approach, we integrate both systems on the same hardware and share some ROS packages. The goal of achieving complete independence remains part of the project’s objectives, and we believe that, in principle, this architecture allows the two systems to be separated. Ideally, different computers or processors would be used, with redundant sensors and components, ensuring no integration between them.

In this work, the SRAS and SS share the mobile platform, LiDAR percepts, and computing resources, but it is important to note that some independence is still maintained: the \gwendolen\ agent implementing the SS runs outside ROS in a \java\ BDI framework (MCAPL), and the SRAS control system uses Python and C++, providing software diversity and interface segregation. Integration is managed through \texttt{java\_rosbridge} and an orchestrator node, which ensures that the SS can act over the SRAS, taking control when necessary and enforcing its priority, e.g., by blocking or overriding SRAS signals. In this section, we present how this integration has been achieved.

\begin{figure}[t]
    \centerline{\includegraphics[scale=0.284]{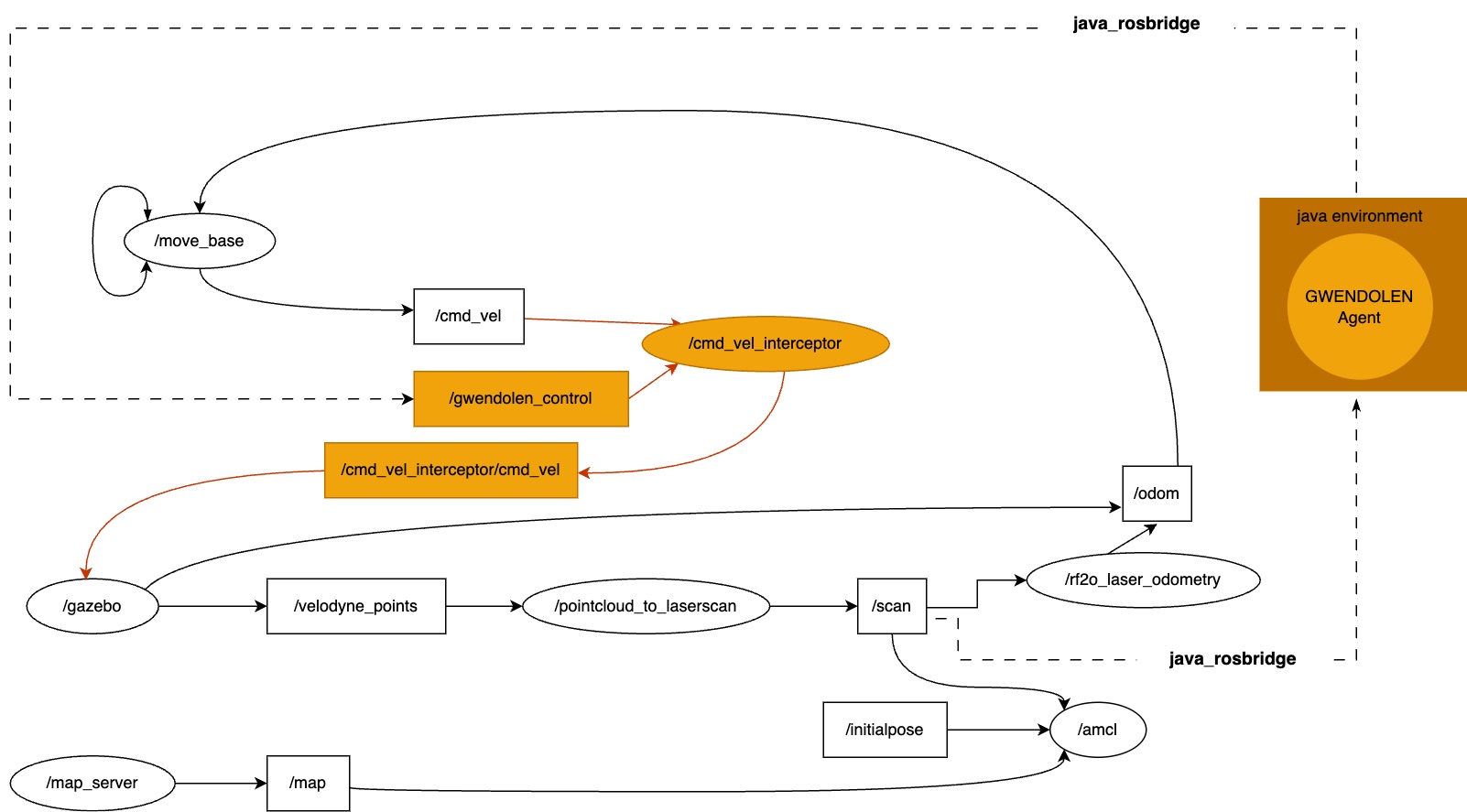}}
    \caption{Integration of SRAS with the SS (\gwendolen\ BDI Agent) on the same hardware. Sharing mobile platform, LiDAR percepts, and computational resources, and communicating via \texttt{java\_rosbridge}.}
    \label{Fig:SRAS_SS}
\end{figure}

\textbf{Figure}~\ref{Fig:SRAS_SS} illustrates the integration of SRAS and SS. The light orange circle highlights the \gwendolen\ agent, which runs in a separate environment implemented in \java. The agent uses \texttt{java\_rosbri{\-}dge} to publish messages to the \texttt{/gwendolen\_control} ROS topic whenever a safety violation is detected. On the ROS side, we developed a custom Python node, called \texttt{cmd\_vel\_interceptor}, which acts as an orchestrator between the autonomous navigation system and the safety subsystem. This node subscribes to both the \texttt{/gwendolen\_control} topic and the standard \texttt{/cmd\_vel} topic generated by \texttt{move\_base}. Under normal conditions, the \texttt{cmd\_vel\_interceptor} node transparently forwards velocity commands from \texttt{move\_base} to the robot’s actuators. However, if a \texttt{true} message is received on the \texttt{/gwendolen\_control} topic, the node overrides all incoming velocity commands and instead publishes a zero-velocity \texttt{Twist} message, stopping the robot. When the safety condition is cleared (i.e., no new stop signals are received), normal velocity forwarding resumes.

This approach ensures that the Safety System (SS) can always enforce safety constraints independently of the SRAS decision-making. It allows the SS to intervene at any time, providing an isolated layer of safety supervision without interfering with the normal operation unless a hazard is detected.

\subsection{Verification and Validation} 
\label{sec:verification}

To demonstrate the reliability of our approach, it is important to provide strong evidence that the system meets its requirements. This involves formal verification and validation. For verification, we use model checking and an automatic correctness proof, ensuring that the system behaves as expected. Validation is carried out through testing, allowing us to observe system performance under realistic conditions. Similar corroborative verification and validation strategies have been successfully applied in~\cite{webster2020corroborative}, ensuring reliability, trustworthiness, and safety in autonomous systems. In this subsection, we present the verification and validation process, beginning with the elicitation of the safety requirement.

\subsubsection*{Formal Requirement Elicitation} 
\label{sec:FormalRequirementElicitation}

\begin{figure}[t]
\centering
\includegraphics[width=1\linewidth]{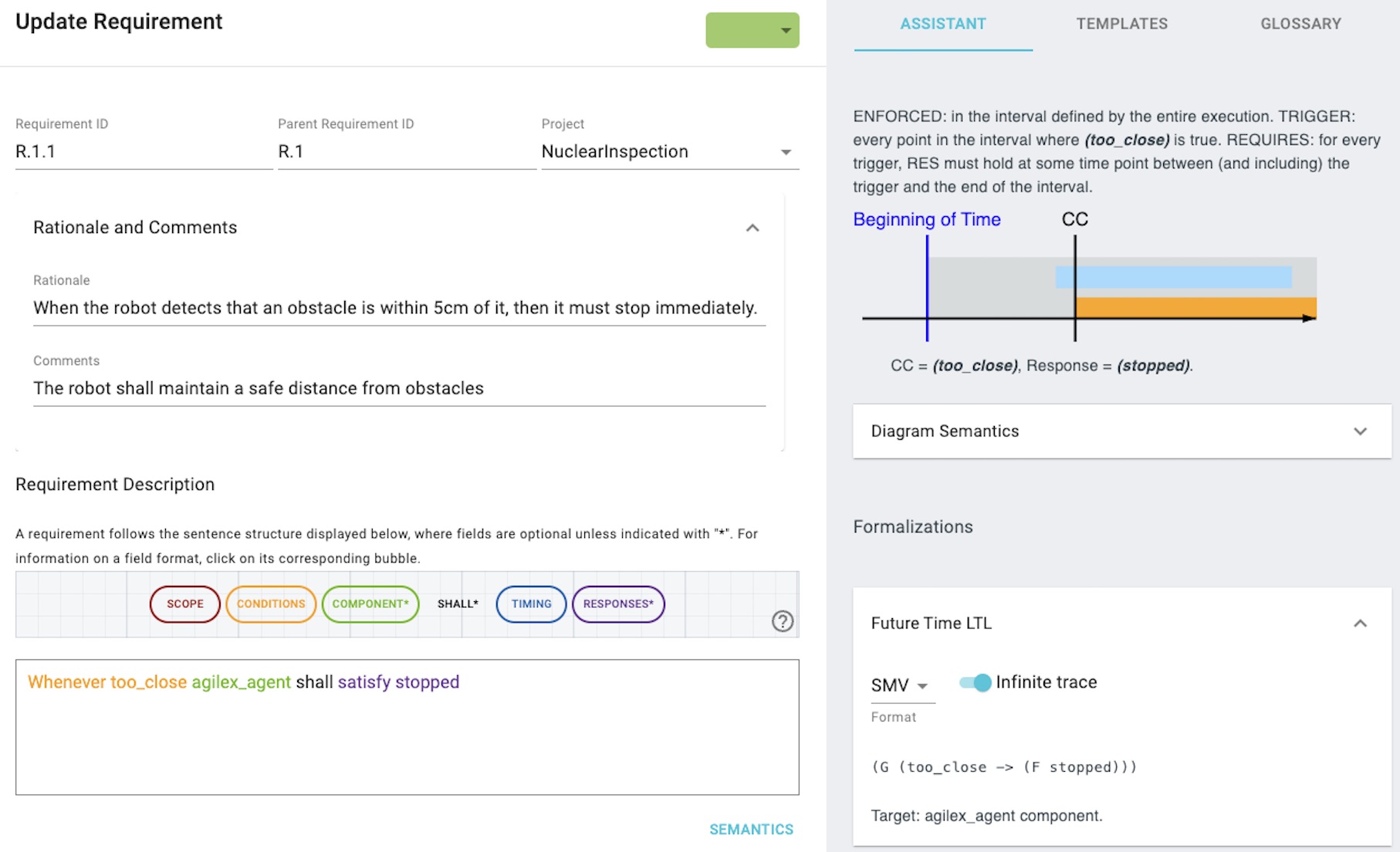}
\caption{Requirement in FRET. We use FRET to translate the requirement from natural language into Linear Temporal Logic (LTL). For model checking, we use the Future Time interpretation. }
\label{fig:FRET-requirement}
\end{figure}

From \textbf{Section}~\ref{Sec:SafetyInstrumentedFunction}, we already know that the Safety System has been programmed to maintain a safe distance from obstacles. However, to formally verify the implementation of the SS, we need to express this requirement as a formal logical property. This is necessary to apply model checking, which is the primary verification technique used in this work.

For this purpose, we use the Formal Requirements Elicitation Tool (FRET)~\cite{pressburger2022fret}, which enables us to translate requirements from natural language into formal specifications, such as Linear Temporal Logic (LTL). \textbf{Figure}~\ref{fig:FRET-requirement} shows the FRET interface used to formalize the safety requirement. This requirement is written using a structured natural language called FRETish, which allows us to interact effectively with the tool. FRETish comprises the following sequential fields: 

\centerline{\fretishComponents}

The component and response fields are compulsory, along with the ``shall'' keyword. FRETish requirements describe the scope and conditions under which a defined component must satisfy a specified (potentially probabilistic) temporal response. This structured format facilitates the translation of natural-language requirements into a form that FRET can process. By default, when certain fields are omitted, FRET assumes sensible defaults: if the scope is omitted, it is taken as global (meaning the property is expected to hold during all possible modes of operation); if the condition is omitted, it is interpreted as true; and if the timing is omitted, it defaults to eventually, to specify that the response is required to hold at least once between the trigger and the end of the interval~\cite{giannakopoulou2021automated}. 

It is worth noting that FRET allows us to write requirements with explicit time dependencies. However, the model checker used in this work (\ajpf) does not support the LTL next operator, and we cannot represent explicit time dependencies. As a result, we interpret the timing field using the default \textit{eventually} semantics, meaning that the agent is expected to stop at some point after detecting a close obstacle. We acknowledge the limitations of this approximation. For more robust verification, a similar approach to~\cite{kamali2017formal}, extracting agent code to check real-time properties, could be applied. This is considered as future work; our current focus is on demonstrating the feasibility of the approach rather than addressing the full technical details of real-time verification.

With these conditions applied, the requirement is instantiated as:

\centerline{
\scope{(global)} 
\conditionF{whenever too\_close}
\component{agilex\_agent} 
shall 
\timing{(eventually)} 
\responseF{stopped}
}

FRET then automatically generates the corresponding Linear Temporal Logic (LTL) formalization\footnote{A PCTL* formalization is generated when the probability field is used.}. In this case, under the Future Time semantics with infinite trace, the LTL expression is: 

$\mathtt{G}\;(\texttt{too\_close} \rightarrow \mathtt{F}\;\texttt{stopped})$

This means that globally, whenever the \texttt{too\_close} condition is true, eventually the \texttt{stopped} condition will hold. The right hand side   of \textbf{Figure}~\ref{fig:FRET-requirement} also includes a diagrammatic semantics of the corresponding FRETish requirement, where the requirement is enforced from the beginning of time and must be satisfied every time that the \texttt{too\_close}  condition holds.

\subsubsection*{SS Verification Using AJPF Model Checking}

\begin{figure}[t]
\begin{lstlisting}[language=java, caption=Verification environment (code snippet), label=lst:verificationEnv, basicstyle=\scriptsize\ttfamily]
...
public class VerificationEnv extends 
VerificationofAutonomousSystemsEnvironment {

    @Override
    public Set<Predicate> generate_percepts() {
        Set<Predicate> beliefs = new HashSet<Predicate>();
        boolean stop_moving = random_bool_generator.nextBoolean();
        if (stop_moving) {
            Predicate stopped = new Predicate("stopped");
            beliefs.add(stopped);
        }
        return beliefs;
    }...

\end{lstlisting}
\end{figure}

\begin{figure}[t]
\begin{lstlisting} [language=prolog, caption=list of verifiable properties, label=lst:psl, basicstyle=\scriptsize\ttfamily]
1a: [] (B(agilex_agent,too_close) -> <> B(agilex_agent,stopped))

% [] means "always" (globally)
% <> means "eventually"
% B(agent, proposition) means "agent believes proposition"
\end{lstlisting}
\end{figure}

\label{sec:SSVerificationAJPF} We created a \emph{verification environment}, shown in \textbf{Listing}~\ref{lst:verificationEnv}, in which to model check our \gwendolen\ program. We abstract away the raw ROS topics, in this case, the LIDAR sensor data, and instead represent the relevant information as percepts or beliefs for the agent. In this way, we focus the verification on the agent’s decision-making logic rather than the complexities and noise of sensor data processing. Further details on verification environments are in~\cite{dennis2023verifiable}, with applications of \ajpf\ in autonomous vehicle decision-making~\cite{alves2019reliable}, ethical robot reasoning~\cite{bremner2019proactive}, and vehicle platooning verification~\cite{kamali2017formal}.

Using the formal safety requirement that was previously specified in FRET as the LTL formula $\mathtt{G}\;(\texttt{too\_close} \rightarrow \mathtt{F}\;\texttt{stopped})$, we translated this into \ajpf’s property specification language as a belief-based property, and added it to the list of verifiable properties, as shown in \textbf{Listing}~\ref{lst:psl}, line 1.

In the verification environment (\textbf{Listing}~\ref{lst:verificationEnv}), we verified that the program satisfies the property: \emph{It is always the case that if the agent believes it is too close to an obstacle, then it eventually believes it has stopped}. The abstraction of ROS topics is implemented in the \texttt{generate\_percepts()} method of the \texttt{VerificationEnv} class (line 6). Here, an initially empty set of beliefs is created (line 7), and a random Boolean value is generated to determine whether the agent should stop (line 8). If the outcome is true, a new predicate \texttt{"stopped"} is created (line 10) and added to the set of beliefs (line 11). Finally, we return the resulting set of percepts (line 13), which represents the agent’s current view of the environment.

\subsubsection*{Deductive Verification Correctness Proof of the Orchestrator Node in Dafny}
\label{sec:CorrectnessProofOrchestratorDafny}

\begin{figure}[t]
\begin{lstlisting}[language=java, caption=Formal model of \texttt{cmd\_vel\_interceptor} in Dafny, label=lst:cmdVelDafny, basicstyle=\scriptsize\ttfamily]
datatype Twist = Twist(x: int, y: int, z: int)
class CmdVelInterceptor {
    var stop_requested: bool    
    constructor ()                          
        ensures stop_requested == false
    {
        stop_requested := false;
    }
    method stop_callback(msg: bool)
        modifies this
        ensures stop_requested == msg
    {
        stop_requested := msg;
    }
    method cmd_vel_callback(msg: Twist) returns (out: Twist)
        ensures stop_requested ==> out == Twist(0, 0, 0)
        ensures !stop_requested ==> out == msg
    {
        if stop_requested {
            out := Twist(0, 0, 0);
        } else {
            out := msg;
        }
    }
}
\end{lstlisting}
\end{figure}

Given the integration of SS and SRAS (\textbf{Subsection}~\ref{Sec:SRASandSSIntegration}), it is crucial to verify that the SS reliably takes priority over SRAS, ensuring the architecture consistently meets safety requirements.

The Orchestrator node (\texttt{cmd\_vel\_interceptor}) is implemented in Python, but for formal verification, we model its core logic in the Dafny programming language, abstracting away ROS-specific details like subscribers and publishers. This approach allows us to focus on verifying the node’s logical behaviour under different conditions. In this implementation, we want to prove that (1) when a stop is requested, only zero-velocity commands are published, and (2) when no stop is requested, input velocity commands are forwarded unchanged.

In our Dafny model, we define a \texttt{Twist} datatype for velocity commands (line 1 of Listing \ref{lst:cmdVelDafny}) and a \texttt{CmdVelInterceptor} class (lines 2-8) with the state (\texttt{stop\_requested}) and two key methods: \texttt{stop\_callback} (lines 9-14) and \texttt{cmd\_vel\_callback} (lines 15-24). These methods include formal contracts that specify expected behaviours using the \texttt{requires} (pre-condition), \texttt{modifies} (frame condition), and \texttt{ensures} (post-condition) keywords. The core part is captured by the \texttt{cmd\_vel\_callback} method (lines 15–25) which returns \texttt{Twist(0,0,0)} if \texttt{stop\_requested} is true, otherwise the input velocity; the \texttt{stop\_callback} method (lines 9–14) updates \texttt{stop\_requested}. In version 3.4.4 of Dafny, we automatically verified three proof obligations (for the \texttt{constructor}, \texttt{stop\_callback}, and \texttt{cmd\_vel\_callback} methods), using version 1.103.1 of VSCode on a Mac M2 Pro running macOS Sequoia 15.6. In this way, we verify that the Orchestrator node satisfies the property: \emph{“If a True stop\_request is received, current and new velocity commands must be replaced with a zero-velocity message”}.

\subsubsection*{Testing} 
\label{Sec:Testing}

We tested the architecture in a Gazebo simulated nuclear waste environment and a physical lab. The experiments included a mission where the robot was programmed to visit three inspection points and then return to its initial position while avoiding obstacles presented in the environment. %This work includes an open-source artifact available via our
%\href{https://github.com/dianabenjumea/Safe-ROS}{GitHub project}\footnote{GitHub project: https://github.com/dianabenjumea/Safe-ROS}.

\begin{figure}[t]
    \centering
    \begin{subfigure}[t]{0.49\textwidth}
        \centering
        \includegraphics[scale=0.098]{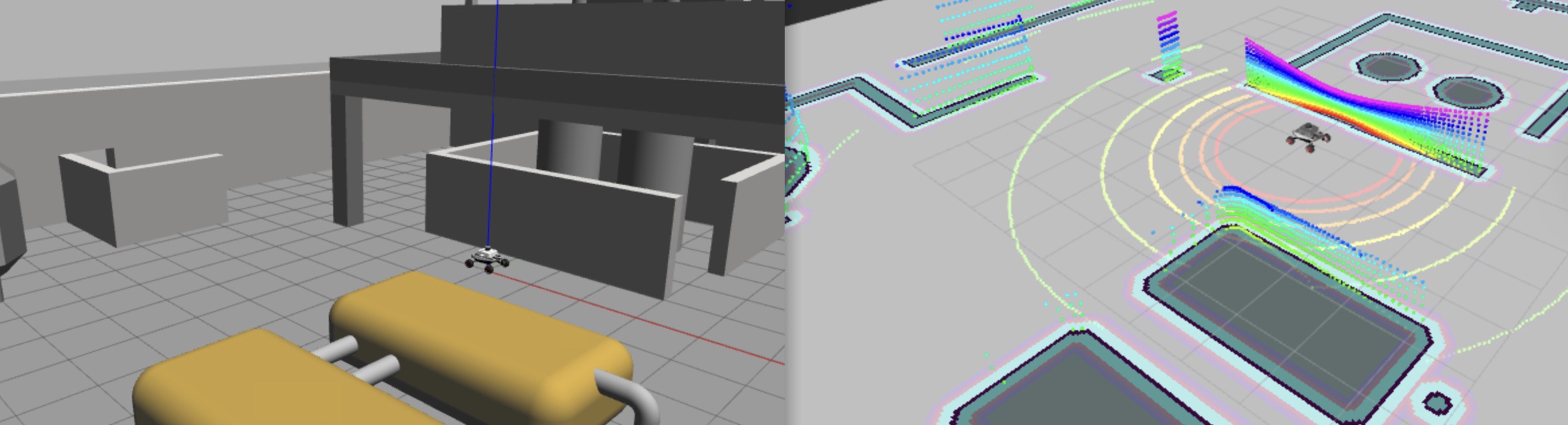}
        \caption{Gazebo simulation using the AgileX Scout Mini platform and a nuclear storage simulation environment}
        \label{fig:gazebo_simulation}
    \end{subfigure}
    \hfill
    \begin{subfigure}[t]{0.49\textwidth}
        \centering
        \includegraphics[scale=0.105]{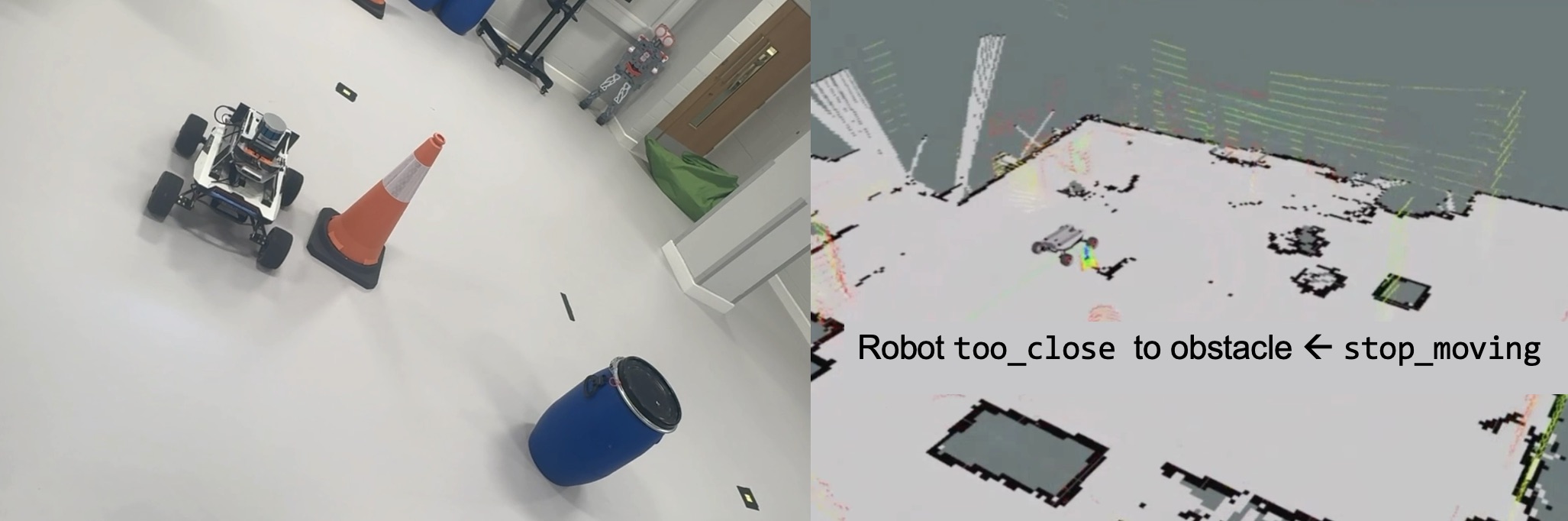}
        \caption{AgileX Scout Mini at the Autonomy and Verification research lab of The University of Manchester}
        \label{fig:lab_experiment}
    \end{subfigure}
    \caption{Simulation and physical testing of the implemented architecture}
    \label{Fig:Combined}
\end{figure}

In the simulation (\textbf{Figure}~\ref{fig:gazebo_simulation}), the environment included walls and nuclear storage elements. The autonomous navigation was run using the ROS \texttt{move\_base} node, and we specifically validated that when the robot approached too close to a wall or object, the \gwendolen\ agent would stop the robot before reaching the safety threshold. These events occurred frequently, which required adjusting \texttt{move\_base} parameters to prevent path computation through unsafe regions. Despite these adjustments, the safety system was triggered multiple times, demonstrating both its effectiveness and its necessity when using probabilistic approaches, such as the algorithms implemented in the \texttt{move\_base} package. It is important to note, however, that stopping the robot does not automatically guarantee a safe state, particularly if the robot stops near a hazardous location. This illustrates the need for a more comprehensive safety argument defining safe states, recovery procedures, and fault tolerance.

In the physical lab (\textbf{Figure}~\ref{fig:lab_experiment}), the environment included cones and barrels that the robot had to avoid. During testing, we deliberately moved some obstacles from their initial mapped positions and ran the autonomous inspection mission. We also moved obstacles dynamically while the robot was operating to trigger the “too-close” perception. In all cases, the robot stopped when detecting an obstacle within the safety threshold, confirming the effectiveness of the safety system under more realistic conditions. This test suite helped us to bridge the reality gap between our static  verification methods (AJPF and Dafny) and actual physical executions. These tests address local runtime safety but do not demonstrate how system-level safety emerges from SRAS–SS interactions, leaving global guarantees as future work.

\section{Evaluation}
\label{Sec:Evaluation}

Our evaluation focused on demonstrating that the proposed approach can provide formal guarantees about the safety of an integrated autonomous inspection system. Using the verification environment and \ajpf\ model checker, we systematically verified that the Safety Instrumented Function behaved according to the formalized requirements derived from FRET, thus directly addressing research question \textbf{Q2} by ensuring that the system meets its safety requirements and behaves as expected.

It is important to note the limitations of this verification approach. Formal requirements that are verified using \ajpf\ do not support explicit timing, which is why the property is specified as \textit{eventually} when we formalized it in FRET (see \textbf{Subsection}~\ref{sec:verification}). Additionally, certain adaptations to the implementation were necessary for reasons of expedience (see \textbf{Subsection}~\ref{Sec:SRASandSSIntegration}). Despite these constraints, the verification effectively demonstrates compliance with the defined safety requirement.

To address \textbf{Q1}, our architecture,  as a framework, can mitigate false results when using probabilistic models at the ROS application layer. %Our SS provides a formally verified layer that monitors application-level behaviours, detecting hazardous situations, and responding appropriately to bring the robot into a safe state
The SS provides a verified layer to monitor application-level behaviours, detect hazards, and ensure the robot reaches a safe state. We integrated the SS and SRAS on the same hardware and shared some ROS packages. However, the approach is designed so that, in principle, the two systems can be separated%, with the SS implementing an independent and diverse SIF
. It is  important to note that our approach is applicable in contexts where safety properties can be explicitly encoded. This is often feasible in nuclear applications, where operations are highly constrained, but may not always be achievable in other domains.

Robotics is an interdisciplinary field, and many components rely on probabilistic methods that improve performance in specific tasks, such as camera image processing and/or path planning. These techniques often involve approximations and cannot guarantee perfect accuracy, which presents challenges for formal verification, because traditional model checking assumes deterministic behaviour. Using our approach, it is not possible to fully verify the probabilistic behaviour of these components. \ajpf\ does not support probabilities, and it is not effective to model functions such as image processing or path planning in \gwendolen\ because a BDI agent abstracts behaviour in terms of beliefs, desires, and intentions rather than probabilistic computations. A fully deterministic model is also infeasible due to variable sensor data. However, by placing these components under the supervision of a formally verified, independent, and diverse SIF, it is still possible to ensure that, even if the underlying components behave probabilistically, critical safety properties are always preserved.

It is also worth noting that the underlying operating system and containerization environment (Docker) used in this work are not formally verified according to Functional Safety (FS) standards~\cite{IEC61508,IEC61513:2011,BS61513}. In practice, a safety-certified operating system could be employed, though this introduces additional complexity, especially in nuclear applications. Alternatively, a safety-verified bare-metal implementation could be adopted, which aligns with the ultimate goals of initiatives such as micro-ROS (e.g., micro\_ros\_arduino), aiming to provide formally verified runtime environments for robotic systems.

Finally, for \textbf{Q3}, we documented the end-to-end process including requirements elicitation in natural language, formalization, implementation, integration, and verification to demonstrate traceability across the system lifecycle. This comprehensive documentation provides evidence supporting the feasibility of our approach to building safe and verifiable autonomous systems in complex robotic environments.

\section{Related Work}
\label{Sec:RelatedWork}

While verification of robotic systems has been considered for many years, it has rapidly grown in attention over the last decade~\cite{ingrand2019recent,luckcuck2019formal}. 
A range of approaches exist of which the most relevant to our work are the use of formal models to verify deliberative components (e.g., the work on the use of verified Cognitive Agents for high-level decision-making in autonomous systems~\cite{dennis2023verifiable,cardoso2020heterogeneous}) upon which we draw and the use of monitors deployed at runtime to constrain the behaviour of the system (e.g., as in ~\cite{mehmed2020monitor}).  Relatively little work has been done in combining the concept of a high-level agent reasoner with that of some kind of monitor, though we note some work in the area of \emph{ethical governors}~\cite{cardoso2021implementing} that adopts this approach, but does not link it to safety engineering processes in the way that we do here.

Verification in robotics generally occurs either at design time or at runtime. Design-time methods (e.g., model checking, theorem proving, or simulation-based validation) verify correctness before deployment, while runtime verification complements them by detecting and mitigating unsafe behaviours during execution~\cite{luckcuck2019formal}. The Safe-ROS architecture facilitates the integration of these methods by using formal verification to prove the correctness of the SIF and deploying it as a runtime safety monitor that supervises the SRAS.

Given the widespread use of ROS, there is a growing interest in adapting it for high-assurance and safety-critical domains, such as aerospace and medicine. For instance, Space ROS~\cite{probe2023space} aims to make ROS suitable for spaceflight applications by improving determinism and reliability, while the ROS-MED project~\cite{rosmed2025} explores its use in medical robotics. These efforts highlight the need for complementary safety assurance mechanisms in regulated environments. However, despite its flexibility, ROS was not originally designed with formal verification or certification in mind. Its use in safety-critical contexts demands additional safeguards such as redundant safety layers or formally verified modules. Our contribution builds on this by integrating a verified agent into a ROS-based architecture, providing a framework for deploying autonomous systems with safety guarantees. This also aligns with emerging standards for autonomy in safety-critical domains. The IEEE 7009-2024 standard~\cite{wallace2025ieee} defines principles for fail-safe design of autonomous and semi-autonomous systems, and ongoing work in the IEEE P7009.1 working group is addressing safety management and interventions during anomalous behaviour\footnote{\url{https://standards.ieee.org/ieee/7009.1/11850/}}.

\section{Discussion and Further Work}
\label{Sec:DiscussionAndFurtherWork}

The evaluation in Section~\ref{Sec:Evaluation} shows that Safe-ROS addresses the research questions in Section~\ref{Sec:Introduction}, while highlighting current limitations and future challenges. For Q1, Safe-ROS demonstrates that an intelligent SS which acts as a safety wrapper can prevent unsafe outcomes by monitoring behaviours and enforcing runtime safety properties, even when the underlying components are probabilistic and/or unverified. However, a full safety argument (defining safe states, recovery procedures, and fault tolerance) is required to substantiate safety claims, and the absence of global or compositional guarantees limits claims about overall system safety. For Q2, formal verification demonstrates that the SIF meets the formalized safety requirements, supporting the correctness of SS logic and its ROS integration, while Q3 demonstrates end-to-end traceability from requirement elicitation to implementation. Together, these results underscore the main contribution of this work: the Safe-ROS architecture, which provides verifiable safety oversight for deploying autonomous robots in safety-critical domains. 

We evaluated Safe-ROS in a nuclear inspection scenario, focusing on a single safety requirement: ensuring the robot stops when too close to an obstacle. The paper documents the full prototype workflow, including motion controller implementation, requirement elicitation, SIF design, SRAS and SS integration, and verification, validation, and evaluation. While this verified property captures only basic safety behaviour, it demonstrates the feasibility of implementing the proposed approach. The current SIF could be implemented using a hardware cut-off, but further work aims to explore more complex behaviour, such as returning to a door, which cannot be achieved with simple hardware guards. It is important to ensure that SRAS communication does not interfere with the SIF’s operation. One approach could involve the SIF using stored waypoints within the SS, which a verified, independent module could access to safely guide the robot back. This functionality constitutes future work.

Our abstraction of ROS topics into agent beliefs enables verification but raises questions about sensor validity and translation correctness. As the evaluation is simulation and lab-based, threats arise from idealized models and lack of real-world testing. We note that the current evaluation establishes only a proof of concept; future work will develop a robust evaluation strategy involving fault injection~\cite{kassem2019detecting}, statistically focused simulation, and experimental campaigns similar to those presented in~\cite[Sections 4--5]{campos2024study}. Physical trials and enhanced sensor modelling are also planned. The tools used (\mcapl, Dafny) would additionally require acceptance by UK nuclear site licensees or regulators before practical deployment, which is beyond the scope of this work.

\textbf{Safety Argument.} While Safe-ROS provides a formally verified SIF that monitors and intervenes on the SRAS, a complete safety argument requires consideration of factors beyond formal verification. Stopping the robot does not always guarantee a safe state, especially if it stops in a hazardous location. Safe states are context-dependent, requiring consideration of the robot’s environment and potential hazards. Our current implementation serves as a proof of concept, demonstrating the feasibility of the approach, but further research is needed to establish that the architecture ensures overall system safety.

The verification performed using \ajpf\ ensures that the SIF behaves correctly according to its formalized internal logic; however, it does not account for perception errors, message delays, or actuator uncertainties. As a result, the safety guarantees apply primarily to the internal decision-making process of the SIF rather than the full, real-world operational system. Recovery mechanisms could be included in the SIF, but their effectiveness depends on the underlying environment and the capabilities of the SRAS.

The architecture does not formally establish how system-level safety emerges from SRAS–SS interaction. The lack of global properties or compositional guarantees limits claims about overall system safety, highlighting areas for future research. Extending Safe-ROS to incorporate global safety reasoning, multi-agent coordination, and richer SRAS models would be necessary to provide stronger assurances in more complex or dynamic scenarios. Despite these limitations, the approach shows that formally verified supervisory control can reduce the risk of unsafe outcomes from probabilistic or unverified components.

\textbf{Scope and Limitations.} While the case study in this paper uses a mobile robot for nuclear inspection, the underlying concept of an independent, verifiable Safety System overseeing an unverified autonomous controller can be extended to other domains (e.g., aerospace, mining, infrastructure) where autonomy is implemented on middleware such as ROS and where external intervention channels (e.g., motion overrides, mode switching) exist. In this implementation, the SS and SRAS were integrated on the same hardware and shared some ROS packages; however, the architecture is designed so that the two systems could be fully independent, potentially using redundant sensors and separate hardware to enhance reliability. Safe-ROS is therefore not limited to one application but defines a reusable pattern for modular safety supervision architectures adaptable to other safety-critical contexts.\smallskip

\noindent\textbf{Availability of Materials and Code:} All source code, models, and verification artefacts are open source at \href{https://github.com/dianabenjumea/Safe-ROS}{\texttt{https://github.com/dianabenjumea/Safe-ROS}}
, including the ROS controller, SS, and verification tools, supporting reproducibility and future research.\smallskip

\noindent\textbf{Acknowledgements:} This work was funded in part by The University of Manchester, the EPSRC-funded CRADLE project (EPSRC grant EP/X02489X/1), and the Royal Academy of Engineering, and benefited from a Fellowship at RAICo (The Robotics and AI Collaboration). \smallskip

\noindent\textbf{Competing Interests:} The authors have no competing interests to declare. 

\bibliographystyle{eptcs}
\bibliography{references}

\end{document}